# Ethical Issues with Using Internet of Things Devices in Citizen Science Research: A Scoping Review

James Scheibner, Anna Jobin, and Effy Vayena

*Health Ethics and Policy Lab, ETH Zurich, Switzerland*

**Abstract**

Our chapter presents a scoping review of published scientific studies or case studies of scientific studies that utilise both citizen scientists and Internet of Things devices. Specifically, we selected studies where the authors had included at least a short discussion of the ethical issues encountered during the research process. Having conducted a search of five databases (IEEE Xplore, Scopus, Web of Science, ProQuest, and PubMed), we identified 631 potential results. Following abstract and title screening, and then full text eligibility assessment, we identified 34 published articles that matched our criteria. We then analysed the full text for these articles inductively and deductively, coding ethical issues into three main categories. These categories were autonomy and data privacy, data quality, and intellectual property. We also analysed the full text of these articles to see what strategies researchers took to resolve these ethical issues, as well as any legal implications raised. Following this analysis, our discussion provides recommendations for researchers who wish to integrate citizen scientists and Internet of Things devices into their research. First, all citizen science projects should integrate a data privacy protocol to protect the confidentiality of participants. Secondly, scientific researchers should consider any potential issues of data quality, including whether compromises might be required, before establishing a project. Finally, all intellectual property issues should be clarified both at the start of the project and during its lifecycle. Researchers should also consider any ethical issues that might flow from the use of commercially available Internet of Things devices for research.

James Scheibner, Department of Health Sciences and Technology, ETH Zurich; Anna Jobin, Department of Health Sciences and Technology, ETH Zurich; Effy Vayena, Department of Health Sciences and Technology, ETH Zurich.

Correspondence concerning this paper should be addressed to Effy Vayena, Department of Health Sciences and Technology, ETH Zurich, Switzerland. Email: effy.vayena@hest.ethz.ch.

**Introduction**

Our chapter seeks to address the ethical issues arising from a collision between two trends in scientific research. First, an increasing amount of research is being carried out by non-professional scientists cooperating with professional scientists (Cooper, 2016; Isling 2018). Secondly, because of the rise in portable and networked computers (henceforth referred to as "Internet of Things"), researchers now have low cost data gathering devices at their disposal. The widespread availability of Internet of Things tools increases the capacity of researchers to collect and process enormous amounts of data (Rothstein et al, 2015; Auffray et al, 2016). Yet scientific projects involving citizen participants may carry a number of ethical complications, including those that may not be immediately apparent to the research team (Cooper et al, 2019). These ethical considerations may be further exacerbated by the ubiquity and massive data gathering potential of Internet of Things devices. However, it is unclear how ethical issues arising in such projects are addressed in practice, and whether they are addressed at all. A brief literature research of published studies did not reveal any review of ethical issues in citizen science related to the use of Internet of Things devices.

We therefore conducted a scoping review of the literature. Its purpose was to analyse whether, and how, ethical challenges for citizen science research involving Internet of Things devices are reported and handled. We aimed at identifying whether researchers in the field are reporting ethical issues and, if yes, what strategies they use to resolve them and what legal implications they mention. Accordingly, our chapter is split into three sections. The first part centres on our methodology and describes the scoping review protocol that was used to identify relevant sections of the literature. The second part offers an analysis of the results that address ethical issues in studies combining citizen science and Internet of Things devices. The third part discusses these results in conjunction with existing theoretical frameworks designed to help guide citizen science projects, and offers recommendations for future research.

**Part 1: Scoping Review Protocol**

In spring 2020, we designed and conducted a scoping review with the goal of retrieving and identifying scholarly literature of studies at the intersection of citizen science and Internet of Things that mention ethical issues. We endeavoured to include articles describing or discussing an empirical study or project involving citizen science and Internet of Things devices, even if they may be using a different nomenclature. We designed and carried out a scoping review by retrieving potentially relevant literature, selecting eligible articles and analysing the relevant sections (Arksey & O'Malley, 2009).

**Retrieval**

Based on our research question we defined the following three relevant root keywords: "citizen science", "ethics", and "Internet of Things". From these root keywords, we then generated a number of alternative, synonymous keywords that includes terms that are more likely to be used in applied literature (cf. Table 1).

TABLE 1

| Keyword | citizen science | ethics | Internet of Things |
|---|---|---|---|
| Alternative terms | citizen science | ethic* | Internet of Things |
| | citizen participation | IRB | IoR |
| | | | Wearable |
| | | | Web of Things |
| | | | mobile device |
| | | | Internet connected |
| | | | Connected device |
| | | | Ubiquitous computing |
| | | | Pervasive computing |
| | | | Smartphone |
| | | | Smart device |
| | | | Sensor |

We searched the following five databases to search for relevant articles: IEEEXplore, ACM Digital Library, Scopus, Web of Science, and PubMed (cf. Table 2). We created the following search strings for each database:

TABLE 2

| Database | Search | Results |
|---|---|---|
| IEEE | ("Citizen science" AND ethic* AND ("Internet of Things" OR "IoT" OR "Internet of Services" OR "Wearable" OR "Web of Things" OR "mobile device" OR "Internet Connected" OR "Connected Device" OR "Ubiquitous Computing" OR "pervasive computing" OR "Smartphone" OR "Smart device" OR "Sensor")) | 9 results |

| ACM Digital Library | ("Citizen science" OR "Citizen participation") AND (ethic* OR IRB) AND ("Internet of Things" OR "IoT" OR "Internet of Services" OR "Wearable" OR "Web of Things" OR "mobile device" OR "Internet Connected" OR "Connected Device" OR "Ubiquitous Computing" OR "pervasive computing" OR "Smartphone" OR "Smart device" OR "sensor") | 122 results |
|---|---|---|
| Scopus | ALL ("citizen science") AND ALL(ethic* OR irb) AND ALL("Internet of Things" OR "IoT" OR "Internet of Services" OR "Wearable" OR "Web of Things" OR "mobile device" OR "Internet Connected" OR "Connected Device" OR "Ubiquitous Computing" OR "pervasive computing" OR "Smartphone" OR "Smart device" OR "sensor" ) | 455 results |
| Web of Science | ALL=(citizen science OR citizen participation) AND ALL=(ethic* OR IRB) AND ALL=("Internet of Things" OR "IoT" OR "Internet of Services" OR "Wearable" OR "Web of Things" OR "mobile device" OR "Internet Connected" OR "Connected Device" OR "Ubiquitous Computing" OR "pervasive computing" OR "Smartphone" OR "Smart device" OR "sensor") | 36 results |
| PubMed | ((("citizen science" OR "citizen participation")) AND (ethic* OR IRB)) AND (internet of things OR IoT OR internet of services OR wearable OR web of things OR mobile device OR connected device OR ubiquitous computing OR pervasive computing OR smartphone OR smart device OR sensor) | 9 results |
| **Total** | | 631 |

We counted 631 matches in total, resulting in 608 articles once duplicates removed. Each of these results was screened by manually examining the title and abstract using the criteria for inclusion and exclusion described in table 3. The inclusion criteria were not applied automatically (that is, we did not search to see whether the text contained the words "citizen science" or "citizen participation" and exclude only using that in the abstract). For example, a project that described volunteer collaborators was not removed because it simply did not contain a mention of citizen science in the abstract. Instead, we manually read each of the titles and abstracts to see whether they matched our screening in or screening out criteria.

**Selection and eligibility**

TABLE 3

| Exclusion Criteria | Inclusion Criteria |
|---|---|
| In the Title or Abstract:<br>● No mention of a study involving citizen participation or citizen science or any synonymous activity (using the search criteria we had developed above) OR<br>● No mention of internet of things, wearables or other synonymous devices (using the search criteria we had developed above) | In the Title or Abstract:<br>● Describing the enrolment or inclusion of citizens or public participation in a scientific project (this can include synonyms for citizen science, such as "public engagement", "crowdsourcing" or "volunteer project") AND<br>● Describing the use of Internet of Things technology in this citizen science-based study or using a synonymous term from the search criteria above (such as mobile devices, sensors, smartphones, and wearables) |

The authors then worked together to assess whether the list of records that they had prepared were congruent with one another and achieved mutual agreement through reflective equilibrium (Daniels, 1996). This resulted in 133 articles screened in, of which we retrieved the full text and proceeded to the eligibility assessment (cf. table 4).

TABLE 4

| | Exclusion Criteria | Inclusion Criteria |
|---|---|---|
| Eligibility Phase | ● One of the following study designs:<br>   o Systematic or scoping reviews<br>   o Policy or meta-analysis articles attempt- | ● One of the following study designs:<br>   o A research study report<br>   o A research study protocol<br>   o A case study or multiple case studies of a citizen |

| | | | |
|---|---|---|---|
| | | ing to design an ethical framework for using citizen science<br>● In the full text of the article:<br>  o Only tangential discussion of citizen science (for example, in journal article title in a bibliography) OR<br>  o Only tangential discussion of ethics OR<br>  o Only tangential discussion of Internet of Things or one of the synonyms included above OR<br>  o Only cursory discussion of ethics approval or ethical issues | science project involving Internet of Things devices.<br>● In the full text of the article:<br>  o A substantive discussion of citizen science, such as in the context of a research project AND<br>  o A substantive discussion of the ethical issues involved in establishing a citizen science project AND<br>  o A substantive discussion of either Internet of Things technology or one of the synonyms included above in the search terms | |

For the eligibility criteria defined above, a substantive discussion includes everything beyond a simple mention of an issue's existence. Even short paragraphs were included to be as expansive as possible with the search criteria (Crampton et al, 2016). To this end, we included all articles as eligible that described a specific study design involving active participants. In contrast, we did not include study designs where the sole involvement of citizens consisted of them passively contributing data about themselves as part of a survey. We also included articles that described case studies, or synthesised a research protocol from existing studies.

After full text eligibility assessment, a total of 34 articles were included as part of the full text analysis. These articles were published across a range of fields between the years 2009 and 2020.

We coded all articles inductively and deductively, identifying ethical issues which we then grouped into clusters for an in-depth analysis.

## Part 2: Analysis of Results

In this section we address the legal and ethical factors raised by the articles we included in our study. We identified the occurrence of three overarching categories of ethical issues: participant autonomy and privacy, data quality, and intellectual property and labour. We will discuss each of these in detail below.

### Participant autonomy and privacy

Existing ethical frameworks require scientific researchers to guarantee the autonomy and safety of all participants in research. This is usually expressed by the default requirements for researchers to seek explicit, informed and free consent from participants prior to research. A number of results in our sample explicitly addressed this question or sought to guarantee participant consent (Seitzinger et al, 2019 (a); Seitzinger et al, 2019 (b); Sousa et al, 2020). For example, Denefleh, in using a sensor device for measuring consumption in a share house, considered whether consent would be affected by the need for housing (Denefleh et al 2019). Likewise, English et al. discuss the importance of ensuring that citizen science studies do not "fall through the cracks" and avoid ethics review or the need for consent (English et al, 2018). It is also important to recall that much of the existing ethics frameworks for scientific research, such as the Nuremberg Code and the Belmont Report, were developed following unethical and harmful research involving minority populations. Therefore, it is important that scientific researchers working with citizen scientists from minority communities avoid repeating the errors of the past. In particular, Pejovic and Skarlatidou highlight the importance of obtaining free, prior and informed consent when working with indigenous populations. This consent includes a requirement that not only should consent be obtained, but the research goals are conveyed to the community (Pejovic & Skarlatidou, 2020).

Unless the participant has expressly indicated otherwise, it is also important to ensure that the confidentiality of participants is protected. Therefore, a number of studies in our sample defined strategies in order to maintain participant privacy, including anonymising or encrypting participant data (Acer et al, 2019; Guerrero et al, 2016; Katapally et al, Komninos et al, 2019). As an alternative but complementary strategy, some studies recommended the use of aggregate data. By using aggregate data, the scientific researchers ensured that individual participants could not be re-identified from their contributions. Nevertheless, statistical disclosure controls should be used following the release of data to protect against re identification from inference attacks (Havinga et al, 2020). Finally, Drosatos et al and Havlik et al describe specific algorithmic platforms to allow citizen scientists to participate in research. These platforms rely on novel privacy enhancing technologies, such as homomorphic encryption, to protect the identity of participants included in research (Drosatos et al, 2014; Havlik et al, 2013).

Some studies reported excluding some forms of participant data where it was judged to be an inappropriate encroachment upon participant privacy. For example, in Acer et al, the research team supplied Belgian postal workers with Android Wear devices to track their movements upon their rounds. However, these devices not only captured geolocational data but also audio data, which the authors acknowledged represented a privacy concern for both the postal workers and their customers. Therefore, as their study was part of a pilot project, the authors determined to disable this continuous audio sensing functionality as part of future research projects (Acer et al, 2019). Conversely, it may not be possible to obtain explicit consent for all forms of data, such as crowd sourced or volunteered geographic information, or social media data. Havinga et al suggest that researchers establishing citizen science projects consider whether mechanisms such as geotagging opt in on a social media platform, represents adequate consent (Havinga et al, 2020).

Another issue related to privacy and raised by Sousa et al is the question of return of results. Several legislative data protection and privacy frameworks provide individuals with the capacity to request data about themselves. In discussing the results of participants collecting data via smartphones from mosquito traps, the authors suggest participants should have the capacity to request data about their contributions (Sousa et al, 2020). Likewise, Katapally et al provide functionality to allow scientific research participants to exercise their right to withdraw from a smartphone based public mHealth study (Katapally et al, 2018). Finally, two of our results, in providing a series of case studies of citizen science projects defined specific protocols for dealing with sensitive data. These sensitive forms of data can include political opinions or the identity of park rangers investigating controversial ecological issues such as cattle invasions or poaching (Heiss & Matthes, 2017; Pejovic & Skarlatidou, 2020). In a similar fashion, Acer et al note the importance of ensuring that activity data from workers will not be used against them by their employer (Acer et al, 2019).

Some of the studies included in our sample also addressed the more abstract question of autonomy, agency and why citizen scientists participate in research. Vesnic-Alujevic et al note that citizen scientists recruited for experiments designed to fine tune wearables for health monitoring are also personalising and actively engaging in their healthcare (Vesnic-Alujevic et al, 2018).

Likewise, Seitzinger et al report how an app for patients to self-report data on foodborne illness study allowed for more sensitive forms of data collection (such as information on milder illness). Further, the authors describe how this approach helped them to avoid complicating factors around privacy and security for the volume of data usually accompanying big data research (Seitzinger et al, 2019(a)).

**Data quality and integrity of citizen science research**

Another fundamental principle of scientific ethics pertains to the quality and integrity of research. In particular, a growing movement has focused on ensuring scientific researchers using computational tools and big data methods can guarantee the verifiability and reproducibility of research (Stodden et al, 2016). However, some of the strongest motivators for citizen scientists to participate in research include personal interest and political reasons. Opponents to citizen science argue that the vested interests of citizen scientists may undermine the accuracy and reliability of the data they contribute. In a similar fashion, a number of the studies included in our sample reported discarding or questioning data due to data quality issues (Andersson & Sternberg, 2016; Aoki et al, 2009; Barzyk et al, 2018; Theunis et al, 2017; Vesnic-Alujevic et al, 2018). The nature of volunteered geographic or crowdsourced information means there can be substantial variances in data quality that are difficulty to calibrate in the laboratory (Elwood et al, 2012; Ferster et al, 2013; Havlik et al, 2013; Komninos, 2019; Weir et al, 2019; Wiggins & He, 2016; Wylie et al, 2014).

However, in our sample we also observed a number of strategies to resolve these issues and guarantee the quality of data. For example, Black and White, as part of an interview study with individuals who contribute air quality readings, note that researchers should consider the implications of "data empowered global citizens". Black and White then report on how interviewees pondered whether they would decide to move from a particularly polluted area if they suffer from respiratory diseases (Black & White, 2016). Another example is the question of how government policy and government-citizen relations may be influenced by citizen science studies. Carton and Ache note that despite criticisms about data quality undermining the integrity of citizen science, citizen sensor networks provide residents with increased "information power" to confront governments (Carton & Ache, 2017). To legitimise this feedback between governments and citizens, Barzyk et al recommend that government agencies publish guidelines on data quality (Barzyk et al, 2018). Some studies already integrated government standards for data quality into their reporting. Aoki et al note that in the context of air quality data, California's *Clear Air Act 1967* creates the regulatory framework for air pollution management and standards.

Related to issues about the political nature of data are concerns regarding data bias. Acer et al note that a majority of data contributions are made by a minority of contributors, which can increase the unrepresentative nature of the sample (Acer et al, 2019). Conversely, the availability of Internet of Things devices may be comparatively less amongst older, regional, and minority populations, leading to a demographic skew in participants (Havinga et al, 2020). Likewise, in Yu et al an entire study was built around addressing deficiencies in data about socioeconomic features of agricultural land systems (Yu et al, 2017). Bias may also be an inherent feature of the data itself, or even exist with the scientific research team processing the data. Heiss and Matthes note that data bias is a particular problem for qualitative social sciences research data, which is based on human perception (Heiss & Matthes, 2017). For crowdsourced data, Wiggins and He note that data from contributors who have previously donated high-quality data may be prioritised over other sources (Wiggins & He, 2016).

In addition to individual and systematic bias, there may be data quality issues associated with the devices used to collect data. In describing how low-cost smartphones and wearables can be used to collect air quality data, Theunis et al point out strategies that can be used to enhance the usability of this data. These strategies can include charging the battery of the measuring device or turning off the measuring software after use. Further, Theunis et al describe how more of these measuring errors arise during the later stages of the project, possibly due to decreasing participant motivation (Theunis et al, 2017). Drawing on the literature from human computer interaction, Budde et al describe how rewards, similar to those used for computer games, can increase participant motivation and guarantee data quality (Budde et al, 2016).

Conversely, the authors in some of our studies recognised that stringent technical standards of data quality could undermine the purposes of the study. To this end, Aoki et al report that in assessing air quality, less accurate but cheaper data collection methods could provide useful information on dramatic regional variances in pollution (Aoki et al, 2009). Likewise, Dema et al suggest that rather than focussing on study protocols, other strategies could be used to improve data quality. These include using tools that collect longitudinal data, as well as more closely integrating participants into the research protocol (Dema et al, 2019). Further, Ferster et al and Heiss and Matthews both note that data quality can be improved through suitable training for volunteers and through focussing on particular areas (Ferster et al, 2013; Heiss & Matthes, 2017). Finally, Drosatos et al note that privacy enhancing technologies for preserving participant confidentiality may necessitate compromising on data quality (Drosatos et al, 2014).

**Intellectual property, data rights and confidential information**

Intellectual property and data ownership may refer to a number of overlapping rights. Each of these rights may apply to different aspects of citizen science research driven by Internet of Things devices. First, there is the question of data control, particularly for open data generated using citizen science projects. A prevailing ethos in citizen science research is the importance of openly available data (Weir et al, 2019). In particular, Komninos reports that ensuring data was made openly available was an incentive for citizen scientists to participate in the project (Komninos, 2019). Further, a number of the studies included in our sample described the benefits of using low cost open access technologies for ubiquitous research (Black and White, 2016; Carton & Ache, 2017). Guaranteeing privacy for participants and ensuring data quality, particularly for the reproducibility of research, represent two competing considerations militating against the use of open data (compare Denefleh et al, 2019 to Drosatos et al, 2014).

However, the presence of intellectual property and moral rights over data can also influence whether that data is made openly available. Unfortunately, the lack of information in this area can present a challenge for researchers planning to use both open data and open source technology. For example, Wylie et al describe how a collective for environmental citizen science encouraged the hosting research institute to update their policies on licensing for open source technology (Wylie et al, 2014). Nevertheless, these issues must be resolved on a case by case basis. Verma et al report on how the ownership of data and images about wildlife could not be transferred across borders due to the potential of identifying endangered species (Verma et al, 2016). Likewise, Yu et al note that the ethics of crowdsourcing big data from farmers as part of agricultural research may depend on who is collecting this data. In particular, industrial agricultural businesses such as Monsanto may gain a significant informational advantage over farmers if they freely benefit from such open research (Yu et al, 2017).

An incidental finding to our identification of ethical issues that indirectly relates to intellectual property concerns the type of devices used for research purposes. We find the most frequently used terms to describe tools for citizen science projects were smartphone (n=27), sensor (n=22) and wearable (n=13). We find less than a third of the results included in our sample refer to "Internet of Things" (n=10) as the class of devices used in their research. By contrast, the use of terms associated with customisable devices ("Internet Connected", "Connected Devices", "Ubiquitous Computing" and "Pervasive Computing") is relatively low.

**Part 3: Discussion**

Our scoping review has identified the occurrence of the three overarching categories of ethical issues mentioned in current literature privacy, data quality, and intellectual property. In this section, we will discuss the legal and ethical factors raised by these issues. Moreover, we will offer recommendations on how to construct citizen science projects involving Internet of Things devices that address potential challenges in this regard.

Our research reveals that a number of ethical considerations must be integrated into the project design in a very early stage. Notably, all citizen science projects should have a protocol that adequately protects participant autonomy and privacy. Although privacy and intellectual property are regulated by specific legislation, and have been addressed in other ethical frameworks, these issues are contextually dependent. (Cooper et al., 2019)

The scientific research team should consider whether personal data is being processed as part of the project. In particular, the analysis of many citizen science projects revealed a nebulous distinction between Internet of Things devices that do and do not process personal data. The scientific research team should also consider whether participants may potentially submit sensitive personal data, or whether these data can be inferred about participants. Whether data has been truly anonymised, or could still be considered personally identifying information, depends on both the data and the environment it has been released into. The scientific research team should ensure data privacy by design, and that the Internet of Things devices used by participants are both privacy-enhancing and secure. This security is particularly important in the context of commercially offered smartphones and wearable devices, where the users may not have control over privacy settings. To this end, a commons of resources for ethics with respect to digital medicine and mHealth projects can be crucial for developing contextually appropriate study protocols (Harlow et al, 2020).

Another issue that was included in the theoretical frameworks above, but only briefly addressed in some of our results, was the question of differences in privacy law between jurisdictions. In particular, the recent European Union (EU) General Data Protection Regulation (GDPR) grants data subjects a number of rights over the use of their data. Some of these rights, such as the right to have data transferred from one device to another, may have a direct impact on citizen science with Internet of Things devices. Therefore, researchers should integrate strategies to deal with these concerns in their study protocol. A number of theoretical and case study derived frameworks define how both citizen scientists and Internet of Things devices should be integrated into research projects (Evans, 2020; Rothstein et al, 2015, Quinn, 2018). These frameworks focus on specific ethical and legal issues that may arise from using Internet of Things devices in citizen science projects, including how citizen science projects can comply with privacy legislation in particular jurisdictions. However, our research also identifies ethical issues that may sit outside the realm of a specific field of legislative regulation. These include potential trade-offs between privacy and data quality, which raises further ethical issues and speaks to another important topic we identified.

Crucially, the ethical issues surrounding data quality are dependent on the study design, the discipline and devices in question. To resolve data quality issues as part of citizen science research, researchers must consider a number of factors contextually. Specifically, it is necessary to consider the types of data that are being collected and in what context. For certain types of data such as visual data of wildlife, the accuracy of data might be less important than the portability of devices (Dema et al, 2019; Verma et al, 2016). To this end, it is important to customise or design Internet of Things data collection devices that are appropriate for the environment in which they are used. Pejovic and Skarlatidou observe how a number of citizen science projects involving indigenous populations in regional areas required supplying low cost devices for these communities suited for regional research (Pejovic & Skarlatidou, 2020). Likewise, Younis et al describe how for near field communication (NFC) devices, positioning is vital to ensure the accurate collection of data (Younis et al, 2019). Secondly, it is also necessary to consider alternative strategies to raise data quality and representativeness, as well as reduce bias. In particular, algorithmic strategies to reduce bias may include assigning rewards for less popular or more spatially distributed tasks (Acer et al, 2019). Outside of technical strategies, it may be possible to also crowdsource validating data. This process would involve recruiting a separate set of participants whose task it is to guarantee the validity of data collected by another set of participants (Wiggins and He, 2016). Nevertheless, any strategy to reduce bias should be employed contextually, recognising in some cases respondent bias can offer valuable insights by itself (Havinga et al, 2020).

A final issue that is not addressed by any of the studies included are the legal rights that Internet of Things device developers hold (Montori et al, 2018). This issue is related to the types of devices used for research purposes, as defined by the use of terms above. There are a number of possibilities to explain this finding. A first hypothesis is that terms such as "Internet of Things", "Ubiquitous Computing" and "Pervasive Computing" are academic terms and are not used in a technical context to describe the tools being used. A second one is an inconsistent use of terms across disciplines (Crampton et al, 2016). The third possible explanation is that citizen science research in our sample largely involves smartphones and wearables sold by manufacturers with proprietary clouds ("the intranet of things"; Montori et al, 2018). This third hypothesis is supported by the fact that the majority of the studies (n=27) included in our sample used either apps relying on smartphone sensors or commercially available devices. By contrast, only a minority of studies used custom designed devices, or devices built using microcontrollers such as Raspberry Pi or Arduino circuit boards (Barzyk et al, 2018; Black & White, 2016; Dema et al, 2019; Denefleh et al, 2019; Tironi & Valderrama, 2018; Verma et al, 2016; Wylie et al, 2014).

These commercial devices can be contrasted with custom manufactured open source platforms, which users may require more time to become familiar with (Black & White 2016; Denefleh et al, 2019). In particular, Theunis et al note that no device can be used for pervasive effortless data collection due to cost or inherent quality issues (Theunis et al, 2017). Therefore, the use of commercial devices may represent an appropriate compromise between each of these factors. Nevertheless, proprietary Internet of Things and mobile devices may have security vulnerabilities that may not be revealed to the project team (Montori et al, 2018). These vulnerabilities raise specific privacy concerns for data collectors, as well as concerns about the verifiability of any data collected using these platforms (Schmitz et al, 2018). Further, commercial smartphone and wearable developers may have their own intellectual property rights over data uploaded to their platforms. Therefore, we cannot assume that all open data (including anonymised data) is prima facie ethical. Instead, the decision to use commercial or open source hardware, as well as any intellectual property concerns, should be determined on a case by case basis.

## Conclusion

The increased prominence of citizen science projects has coincided with a proliferation in the number of Internet of Things devices. The portable, low cost and connected nature of these devices has made them ideal for carrying out citizen science research. However, the use of these devices also may raise ethical and legal issues. To identify these legal and ethical issues, we analysed 34 studies from a variety of fields that employed a variety of different citizen science study designs. We identified privacy, data quality and intellectual property related concerns as the three main issues raised by researchers. Building on our analysis of these ethical issues we are able to voice three recommendations for researchers on how they could ethically integrate participants into citizen science research projects. First, researchers should develop a specific protocol for how to ensure both adequate consent and data protection for non-institutional scientific researchers. This protocol should also allow individuals to exercise their rights under data protection or privacy laws (depending on the jurisdiction). Secondly, researchers should consider the types of data that are being collected using citizen science devices, and what the quality requirements for that data are. Thirdly, where possible researchers should consider how intellectual property rights will be handled, and whether these rights might influence the choice of device. Overall, our analysis of these issues contributes to inform future work on specific ethical issues in citizen science research using Internet of Things devices.